# A Res-FCN for Electromagnetic Inversion of High Contrast Scatterers at an Arbitrary Frequency Within a Wide Frequency Band

Hao-Jie Hu, Jiawen Li, Li-Ye Xiao, *Member*, *IEEE*, Yu Cheng, and Qing Huo Liu, *Fellow*, *IEEE*

*Abstract*− **Many successful machine learning methods have been developed for microwave inversion problems. However, so far, their inversion has been performed only at the specifically trained frequencies. To make the machine-learning-based inversion method more generalizability for realistic engineering applications, this work proposes a residual fully convolutional network (Res-FCN) to perform microwave inversion of high contrast scatterers at an arbitrary frequency within a wide frequency band. The proposed Res-FCN combines the advantages of the Res-Net and the fully convolutional network (FCN). Res-FCN consists of an encoder and a decoder: the encoder is employed to extract high-dimensional features from the measured scattered field through the residual frameworks, while the decoder is employed to map from the high-dimensional features extracted by the encoder to the electrical parameter distribution in the inversion region by the up-sample layer and the residual frameworks. Five numerical examples verify that the proposed Res-FCN can achieve good performance in the 2-D microwave inversion problem for high contrast scatterers with anti-noise ability at an arbitrary frequency point within a wide frequency band.**

## I. INTRODUCTION.

MICROWAVE inversion (MWI) can reconstruct the locations and electromagnetic (EM) properties of the target by analyzing the measured scattered field. With the advantages of being nondestructive and noninvasive, it has been widely used in biomedical engineering [1-4], geophysical inspection [5-6], through-wall imaging [7-8], and so on [9-15].

Due to the inherent nonlinear relationship between the scattered field and the EM properties and the ill-posedness of the discretized matrix equations, most microwave inversion methods are based on iterative techniques to minimize the misfit between the measured scattered field data and the reconstructed scattered field data. The common traditional microware inversion methods, such as Born iterative method (BIM) [16-17], distorted BIM (DBIM) [18], variational BIM (VBIM) [19], and contrast source inversion (CSI) [20-22] method, has achieved some good results in MWI. There also some iterative methods are used to solve the variation of the traditional method, such as conjugate-gradient (CG) [23], non-linear conjugate gradient (NLCG) [24], quasi-Newton (QN) [25] and so on. However, the performance of the iterative methods depends on the initial solution which is usually obtained from the linear approximation. Scatterers with large contrasts will bring strong nonlinearity which is hard for the linear approximation to obtain an acceptable initial solution, so the performance of traditional MWI methods will be compromised, and even with a reasonable initial solution, the computational cost will also largely, due to significant cost will be incurred by iterative optimization schemes.

In recent years, with the rapid development of artificial intelligence, machine learning has brought a new way to solve MWI problems, based on its powerful nonlinear mapping ability [26-34]. Among the published machine-learning-based inversion methods, there are two categories that can be classified. The first category is based on the traditional approximation method or the framework of traditional iterative method [26-32]. The advantage of this kind of method is that, compared with traditional iterative methods, the inversion cost can be largely saved after the model is trained, and because of the physical prior knowledge from the traditional methods, this kind of method has a strong generalizability. However, for the high contrast inversion or the large electrical size problems, this kind of method also has to face the similar limitation as the traditional methods. Thus, to solve the inversion problem with high contrasts or the large electrical sizes, the second category based on end-to-end machine learning methods is proposed to map the scattered field data to the EM parameters. For example, a dual module NMM-IEM is proposed to reconstruct scatterers with high contrasts and large electrical sizes [33]. In [34], Simsek proposes several methods to improve the accuracy of one-dimensional high-contrast inversion problem. However, without the aid from the traditional approximation method, the generalizability of this kind of method is a problem. For example, the frequencies of these application are immutable after training, so if the test data comes from other frequencies, then this kind of method need to be re-trained, including re-collecting the training dataset. Moreover, in the real application of inversion, multiple frequency points are often applied due to the different SNR at different frequency points

Manuscript received XX, 2023. This work was supported by Science and Technology Projects of Innovation Laboratory for Sciences and Technologies of Energy Materials of Fujian Province (IKKEM) under Grant HRTP-[2022]-32, and in part by the National Natural Science Foundation of China under Grant 62001406 and Grant 61471105. (Corresponding authors: Qing Huo Liu; Li-Ye Xiao.)

H.-J. Hu, J. Li, L.-Y. Xiao and Y. Cheng are with Electronic Science and Engineering, Institute of Electromagnetics and Acoustics, Xiamen University, Xiamen 361005, China (e-mail: liyexiao16@gmail.com).

Q. H. Liu was with the Department of Electrical and Computer Engineering, Duke University, Durham, NC, USA. He is now with Eastern Institute of Technology, Ningbo 315200, China (e-mail: qhliu@eias.ac.cn).

to make the inversion results more stable; however, multiple frequency points require several trained models to obtain the inversion results, and different frequency points may lead to different training performance. In [32], Li et. al improves the inversion performance by inputting the BP results at multiple frequencies into U-Net, but the algorithm based on BP still cannot avoid the limitation of traditional method. Above all, it is difficult to reconstruct high-contrast scatterers with traditional methods as the core and machine learning as the aid. The method with machine learning as the core can solve the problem of high contrast but lacking generalizability.

Thus, to make the machine learning-based microwave inversion method more generalizable for engineering applications, in this work, a residual fully convolutional network (Res-FCN) is proposed to perform 2-D MWI for high contrast scatterers at an arbitrary frequency within a wide frequency band. By introducing the residual frameworks into fully convolutional network (FCN), the proposed Res-FCN combines the advantages of the Res-Net [35] and FCN [36], which can build a network deep enough to fit the nonlinearity under the combined influence of frequency and the relative permittivity. At the same time, a dataset generation strategy is designed to transfer frequency information implicitly to the network through the dataset. Numerical examples verify that the proposed Res-FCN achieves good performance in the 2-D MWI with high contrast scatterers and has anti-noise ability in an entire frequency band.

Compared with previous research works, the main contributions of this work can be summarized as follows:

(a) Most current machine-learning-based MWI methods only can perform the inversion at the trained frequency. If the frequency of the testing scattered field is changed, the inversion model needs to be retrained, and the corresponding training data also need to be collected with full-wave simulations. This is a time-consuming process for both the training data collection and model training. The proposed Res-FCN can perform inversion at any frequency within a wide frequency band (i.e. 1-2 GHz).

(b) In this work, a new training dataset generation strategy is designed. The frequency sampling interval of the training data in the entire frequency band is 0.005 GHz (a total of 201 frequency points). Meanwhile, in order to fully demonstrate the performance of the proposed Res-FCN in the whole frequency band, the frequency sampling interval of the test data in the entire frequency band is 0.001 GHz (a total of 1001 frequency points). It means that 800 test frequencies are untrained. Thus, the proposed Res-FCN can not only obtain good results at the trained frequency points in the whole frequency band, but also reconstruct the scatterer distribution at untrained frequency points in the whole frequency band.

(c) MWI with high contrasts has always been a challenging task. For the proposed Res-FCN, due to its direct inversion structure, not only good inversion results can be obtained in the entire trained frequency band, but also the scatterers with high-contrast can be well reconstructed.

The rest of this paper is organized as follows. Section II introduces the structure and specific details of the proposed Res-FCN, and then Section III gives four numerical examples to verify the effectiveness of the proposed Res-FCN. Finally, Section IV concludes this work.

## II. RESIDUAL FULLY CONVOLUTION NETWORK

In this section, the structure of Res-FCN is described in detail for the 2-D MWI. The classical structure of the 2-D MWI is shown in Fig. 1. The transmitters and receivers are evenly distributed on a circle centered inside the domain of interest (DoI). In the following, we will introduce the residual framework first and then the residual framework is included into FCN to construct a deep neural network to map the strong nonlinearity of inversion in the entire frequency band. The architecture of the proposed Res-FCN is shown in Fig. 2.

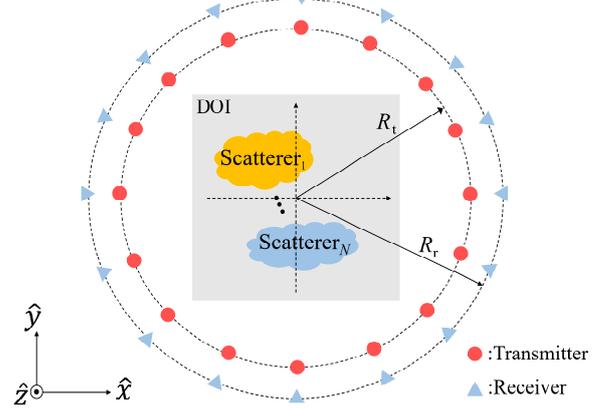

Fig. 1. The classical configuration of a 2-D MWI. Transmitters and receivers surrounding the scatterers are placed on the circles with the radii of $R_t$ and $R_r$, respectively.

### A. Residual framework

To map the scattered field data at different frequencies to the relative permittivity distribution in DoI, three kinds of residual framework are introduced, as shown in Fig. 2 (a)-(c).

For the residual block A, the input and output dimensions of the feature map are set as $x_{\text{dim}} = (L_1, L_2, C)$ and $y_{\text{dim}} = (L_1, L_2, C)$, respectively, where $L_1 \times L_2$ are the dimensions of feature map and $C$ is the channel. The residual block A can be compactly expressed as

$$y = f(x, \{W_i\}) + x \qquad (1)$$

where $y$ is the output of the residual framework, $x$ is the input connected to the output with a shortcut. $f(x, \{W_i\})$, which has two "Conv-BN-LeakyReLU" structures in series, is a nonlinear function with convolution to extract features. In the "Conv-BN-LeakyReLU" structure, the data is input to a convolution layer with convolution kernel size of 3×3, and the convolution stride is set to 1×1 to extract features. Then, a batch normalization (BN) operation is carried out to standardize each dimension of the batches. This operation can accelerate convergence and reduce the influence from the random initialized weights on the network performance [37]. The BN operation can be expressed as

$$y_i = \text{BN}_{\gamma, \beta}(x_i) = \gamma \frac{x_i - \mu_\beta}{\sqrt{\sigma_\beta^2 + \epsilon}} + \beta \qquad (2)$$

where $x_i$ and $y_i$ are the input and output of BN layer, respectively, $\mu_\beta$ is the mean value of $x_i$, $\sigma_\beta$ is the standard deviation of $x_i$, $\epsilon$ is a constant to avoid $\sigma_\beta$ getting close to zero and it is usually set as 0.001, $\gamma$ is the scaling factor, and $\beta$ is the translation bias. $\gamma$ and $\beta$ are the trained parameters in the network.

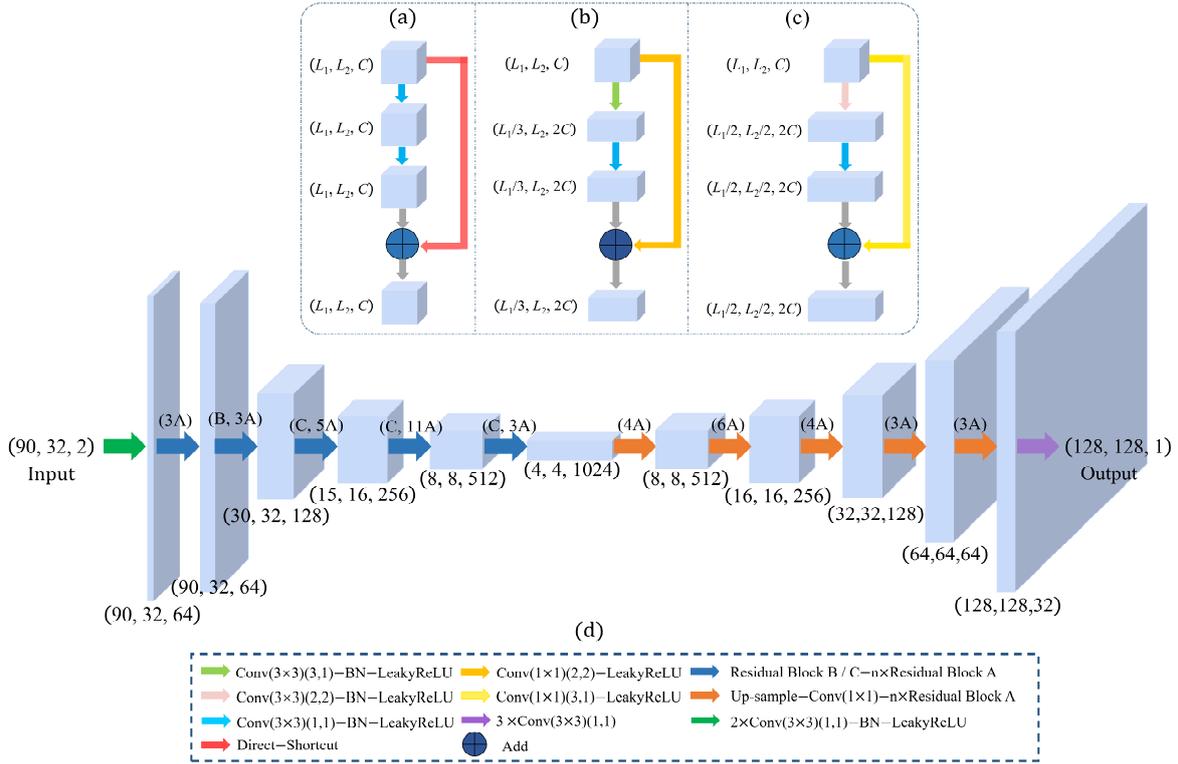

Fig. 2. The network structure diagram of Res-FCN. (a), (b), (c) are three different residual structures, which are called residual blocks A, B and C, respectively. (d) is the main structure of Res-FCN, which is composed of the convolution layer, max-pooling layer, up-sample layer and the residual framework mentioned above. Its input is a two-channel matrix that separates the complex-value scattered field data into real and imaginary parts, and the output is the relative permittivity distribution within DOI.

Then, the output of BN layer is input to the leaky rectified linear unit (LeakyReLU) layer to add nonlinearity mapping ability. LeakyReLU is a variant of rectified linear unit (ReLU) to avoid the output of ReLU getting close to zero. The LeakyReLU layer can be expressed as follows:

$$\text{LeakyReLU}(x) = \begin{cases} x, & x > 0 \\ \alpha x, & x \leq 0 \end{cases} \quad (3)$$

where $x$ is the input of LeakyReLU and $\alpha$ is a small constant, which is set to 0.3.

Except the convolution stride is set to 3×1 and 2×2 of the first convolution in residual blocks B and C, respectively, the residual blocks A, B and C have similar structures. Meanwhile, to avoid losing excessive information while compressing features, the number of channels is multiplied by two in the residual block B and C. Then, the shortcut in the residual block A is changed to a series of convolution layers with the convolution kernel size of 1×1 and activation function LeakyReLU. Meanwhile, to ensure the final addition with the consistent dimensions in the feature map, the convolution stride of this kind of convolution is equal to the first convolution in $f(x, \{W_i\})$.

So, the residual block A is the baseline framework which can be directly used when the input and output have the same dimensions. The residual blocks B and C use convolution stride to replace the pooling layer to complete the feature compression simultaneously. This way can extract information more efficiently during down-sampling, when the network is deep enough [32].

*B. Res-FCN*

Based on the above mentioned three residual frameworks, a deep full convolutional neural network, termed Res-FCN, is proposed to directly map the scattered electrical field to the relative permittivity distribution for the 2-D inversion problem in a frequency band.

Res-FCN consists of two parts, an encoder and a decoder. The encoder module is used to extract and compress the features of the scattered field and map them to high-dimensional feature maps, and then the decoder module maps the obtained high-dimensional feature to the relative permittivity distribution.

As shown in Fig.2(d), the input of the network is the complex-valued scattered field matrix which is separated into the real and imaginary parts and placed in two channels, respectively. It can be compactly expressed as

$$X_k = [\text{Re} \begin{pmatrix} m_{11} & \cdots & m_{1M} \\ \vdots & \ddots & \vdots \\ m_{N1} & \cdots & m_{NM} \end{pmatrix}, \text{Im} \begin{pmatrix} m_{11} & \cdots & m_{1M} \\ \vdots & \ddots & \vdots \\ m_{N1} & \cdots & m_{NM} \end{pmatrix}] \in \mathbf{R}^{N \times M \times 2} \quad (4)$$

where $X_k$ is the $k^{\text{th}}$ training sample, $m_{ij}$ represents the complex-value electrical scattered field data which transmitted by the $i^{\text{th}}$ transmitter and received by the $j^{\text{th}}$ receiver. $N$ and $M$ indicate the number of receivers and transmitters, respectively. The output of the network is the relative permittivity distribution in DOI, which can be discretized and expressed as

$$O_k = \begin{bmatrix} o_{11} & \cdots & o_{1P} \\ \vdots & \ddots & \vdots \\ o_{P1} & \cdots & o_{PP} \end{bmatrix} \in \mathbf{R}^{P \times P} \quad (5)$$

where $O_k$ is the output of network of the $k$th training sample, $o_{ij}$ is the pixel element of $O_k$, and DoI is discretized into $P \times P$ pixels.

There are two parts in the encoder module: the first part contains two convolution layers, a BN layer, and a LeakyReLU layer. In the convolution layer, the size of convolution kernel is 3×3 and convolutional stride is 1×1. After the first part, the preliminary characteristics of the scattered field can be obtained. The second part is used to further extract high-dimensional features from the first part. It is completely composed of the above proposed residual framework. There are one "Residual Block A" structures and four "Residual Block B(C)-Residual Block A" structures, 29 residual frameworks totally; the fourth residual framework is the residual block B, the 8th, 14th and 26th are residual block C, and the rest are residual block A. The penultimate structure is set to have largest number of convolution layers, which is similar to Res-Net [35].

Similar to the encoder, the decoder also consists of two parts. The first part maps the high-dimensional features from the encoder to the feature maps with the same dimension as the electrical parameter distribution. It entirely consists of five concatenated "Up-sample-Conv-Residual Block A" structures, where the numbers of residual blocks in them are 4, 6, 4, 3, and 3, respectively. The up-sample layer is a simple interpolation layer, which is used to double the size of the feature map. Then, a convolution layer with a kernel size of 1×1 and an activation function layer are employed to compress the number of channels to the original half. Finally, the residual block A is used to perform the feature reconstruction on an enlarged feature map. The second part consists of three convolution layers in series. It aims to obtain the final reconstruction result by further compressing the channel. In summary, the decoder module is the inverse process of the encoder, and aims to map the high-dimensional features obtained from the encoder to the distribution of the relative permittivity.

## III. NUMERICAL RESULTS

In this section, five numerical examples are presented to show the inversion performance of the proposed Res-FCN model. To verify the effectiveness of Res-FCN at different untrained frequency points in the entire wide frequency band, the traditional VBIM is employed in the first example as a comparison model. Due to the VBIM restriction, the scatterers in the first numerical example have low contrasts. Thus, the second numerical example is employed to evaluate the inversion performance of the proposed Res-FCN for the scatterers with high contrasts in the entire trained frequency band. The third numerical example is employed to evaluate the anti-noise ability of the proposed Res-FCN. The fourth numerical example is employed to verify the generalization of the proposed Res-FCN. Finally, the experimental test in [39] is used to verify the performance of the model in practical application

All the numerical computation is performed on a computer with an Intel i9-10850K CPU, 128 GB RAM and NVIDIA Geforce RTX 3090 GPU. The spectral element method (SEM) simulation [40] is used to quickly provide training dataset used in this work.

### A. EM Inversion Setups

In this work, the first four numerical examples have an operating frequency band from 1 GHz to 2 GHz, and the corresponding wavelength range in vacuum is from 0.15 m to 0.3 m. DoI is a square with the side length of 0.64 m, and it is discretized into 128×128 pixels where the size of each pixel is 0.005 m×0.005 m. 32 transmitters and 90 receivers are uniformly placed on two concentric circles with the radii of 0.65 m and 0.55 m, respectively. Meanwhile, the interval between adjacent transmitters and receivers are 11.25° and 4°, respectively. In this work, for the 2-D MWI, only $E_z$ of the scattered field data with the dimension of (90, 32, 2) is input to Res-FCN, where the complex $E_z$ is separated into the real and imaginary parts.

In order to quantitatively evaluate the inversion performance of the method, the model misfit and data misfit under $L_2$ norm are defined as follows:

$$\text{Err}_{\text{model}} = \frac{\|\mathbf{M_R} - \mathbf{M_T}\|_2}{\|\mathbf{M_T}\|_2} \qquad (6)$$

$$\text{Err}_{\text{data}} = \frac{\|\mathbf{D_R} - \mathbf{D_T}\|_2}{\|\mathbf{D_T}\|_2} \qquad (7)$$

where $\mathbf{M_T}$ and $\mathbf{M_R}$ are the vectors of the true electrical parameters of the model and their reconstructed values in all the pixels, respectively. $\mathbf{D_T}$ and $\mathbf{D_R}$ are the measured scattered field received by all receivers and the scattered field vector obtained from the reconstructed result, respectively.

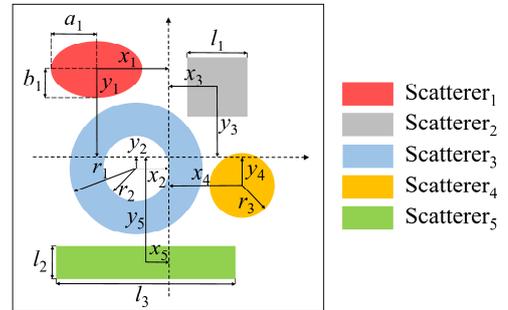

Fig. 3. The composition of the 2-D training model.

### B. Training Details of the Proposed Res-FCN

As shown in Fig. 3, the training model consists of a circular cylinder, a ring, a rectangle, a square and an ellipse, and their positions and sizes are randomly selected in a given range as shown in Table I. $a_1$, $b_1$, $x_1$ and $y_1$ are major axes, minor axes and the centers of ellipse. $r_1$, $r_2$, $x_2$ and $y_2$ are the inner and outer radii and center of the ring. $l_1$, $x_3$ and $y_3$ are the length and center of the square. $r_3$, $x_4$ and $y_4$ are the radii and the centers of circular cylinders. $l_2$, $l_3$, $x_5$ and $y_5$ are set as the length, width and center of rectangle.

Meanwhile, to express the whole information of the wide frequency band, it will require a large number of training samples if the frequency point of each training sample is randomly selected in the range of 1 GHz to 2 GHz. Thus, in this work a new strategy of collecting training samples is proposed. This strategy can be expressed in the following steps:

First, 25000 samples are generated by randomly selecting one to four scatterers shown in Fig. 3, and the corresponding relative permittivity of each selected scatterers are randomly assigned in the range of [1, 8] with the step of 0.1.

Then, one to six frequency points are randomly selected in the frequency band for each sample. In order to fully demonstrate the continuity of field transformation in frequency transformation, the sampling step in the band is set to 0.005 GHz. Thus, there will be one to six sets of scattered field data corresponding to a same ground truth, rather than all the frequency points are employed for a training sample.

Based on this strategy, this work could well capture the information of the frequency band with affordable training cost.

TABLE I
PARAMETER RANGES FOR THE SCATTERERS IN THE TRAINING MODEL SHOWN IN FIG. 3

| Para. | $a_1$ | $b_1$ | $x_1$ | $y_1$ | $r_1$ | $r_2$ | $x_2$ | $y_2$ | $l_1$ | $x_3$ | $y_3$ | $r_3$ | $x_4$ | $y_4$ | $l_2$ | $l_3$ | $x_5$ | $y_5$ |
|---|---|---|---|---|---|---|---|---|---|---|---|---|---|---|---|---|---|---|
| Min. | 0.02 | 0.04 | -0.3 | -0.3 | 0.02 | 0.04 | -0.3 | -0.3 | -0.02 | -0.3 | -0.3 | -0.02 | -0.3 | -0.3 | 0.02 | 0.02 | -0.3 | -0.3 |
| Max. | 0.16 | 0.16 | 0.3 | 0.3 | 0.16 | 0.2 | 0.3 | 0.3 | 0.3 | 0.3 | 0.3 | 0.16 | 0.3 | 0.3 | 0.3 | 0.3 | 0.3 | 0.3 |
| Step | 0.02 | 0.02 | 0.01 | 0.01 | 0.02 | 0.02 | 0.01 | 0.01 | 0.01 | 0.01 | 0.01 | 0.02 | 0.01 | 0.01 | 0.01 | 0.01 | 0.01 | 0.01 |

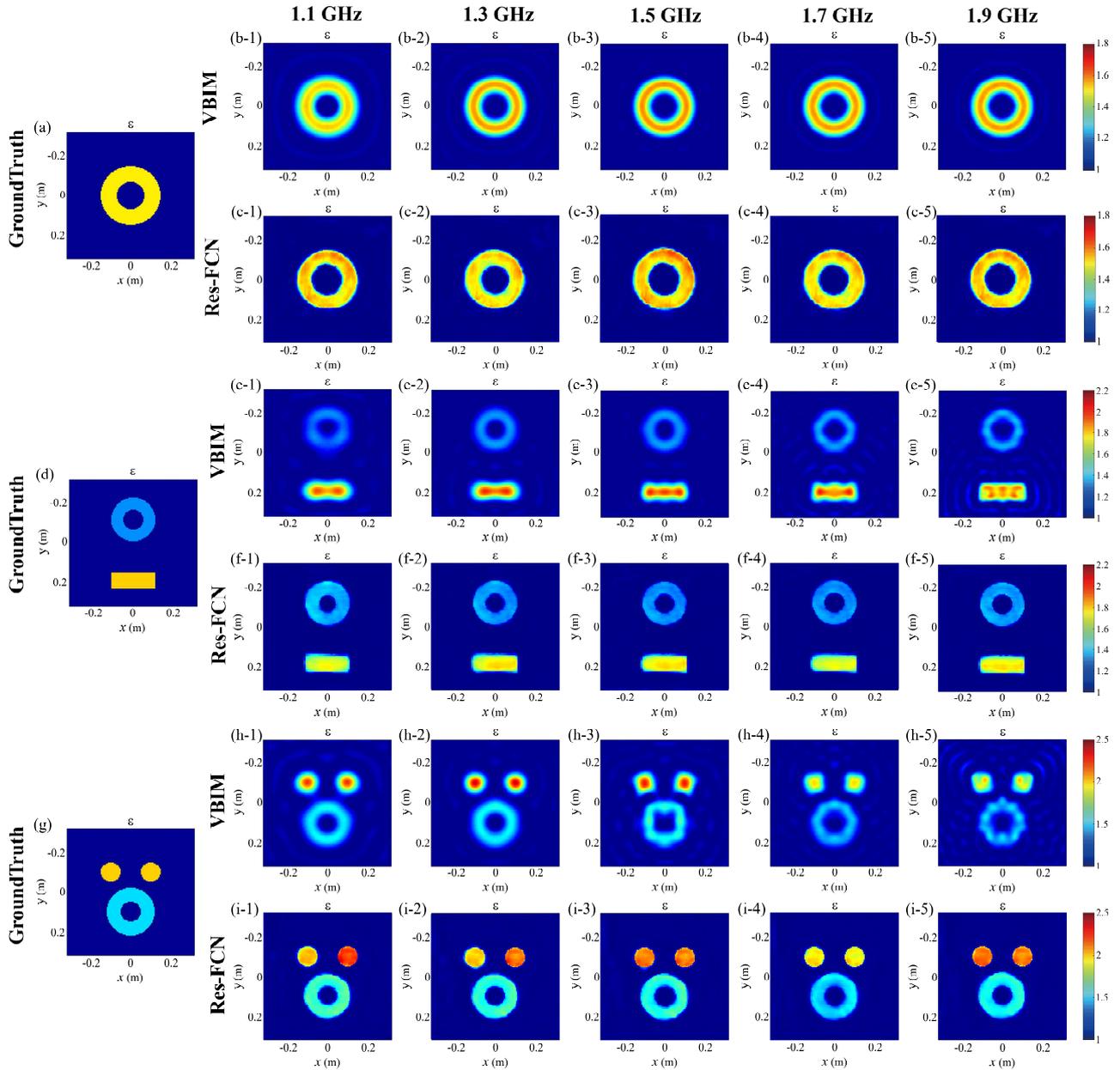

Fig. 4. Three cases are employed to test the proposed Res-FCN and compare with VBIM. (a), (d) and (g) are the ground truth of three cases, respectively. Rows 1, 3, and 5 are the results of VBIM, and rows 2, 4, and 6 are the results of Res-FCN. The results from the first to fifth columns are at 1.1,1.3, 1.5,1.7 and 1.9 GHz, respectively.

In this work, the adaptive moment (Adam) estimation is employed as the optimizer for the network training. The initial learning rate is set to 0.0003. The mean square error (MSE) is used for the loss function of the proposed Res-FCN, and the definition of MSE is as follows

$$\text{MSE}_{\text{loss}} = \frac{1}{N}\sum_{i=1}^{N}(y_i - \hat{y}_i)^2 \tag{8}$$

where $N$ is the number of training samples, $\hat{y}_i$ is the output of the proposed model, $y_i$ is the ground truth. At the same time, in the training process, the data will first pass through a Gaussian noise layer to reduce the over-fitting and make the network more generalizable.

### A. Comparison with Traditional Iterative Method

In order to validate the proposed Res-FCN in the entire frequency band, the traditional iterative method VBIM is employed as the comparison method. In this example, the inversion results of three cases for both methods at five frequency points, i.e., 1.1, 1.3, 1.5, 1.7 and 1.9 GHz uniformly distributed in the operating frequency band, are provided as shown in Fig. 4. Meanwhile, the corresponding model misfits and data misfits are listed in Table II. The value of the relative permittivity of Case #1, Case #2 and Case #3 are gradually increased, and the number of scatterers is also gradually increased from one to three.

| | Frequency (GHz) | 1.1 | 1.3 | 1.5 | 1.7 | 1.9 |
|---|---|---|---|---|---|---|
| Case #1 | VBIM Model misfit (%) | 6.34 | 5.44 | 5.30 | 5.29 | 5.15 |
| | VBIM Data misfit (%) | 0.07 | 0.07 | 0.09 | 0.36 | 0.44 |
| | Res-FCN Model misfit (%) | 2.03 | 2.42 | 1.74 | 1.69 | 2.03 |
| | Res-FCN Data misfit (%) | 4.92 | 8.36 | 4.82 | 4.99 | 5.56 |
| Case #2 | VBIM Model misfit (%) | 7.19 | 6.66 | 6.30 | 6.51 | 6.33 |
| | VBIM Data misfit (%) | 0.10 | 0.09 | 0.10 | 0.14 | 0.22 |
| | Res-FCN Model misfit (%) | 5.43 | 4.19 | 3.91 | 4.34 | 3.84 |
| | Res-FCN Data misfit (%) | 12.01 | 11.93 | 8.97 | 12.91 | 12.55 |
| Case #3 | VBIM Model misfit (%) | 9.41 | 8.70 | 9.24 | 9.38 | 9.98 |
| | VBIM Data misfit (%) | 0.10 | 0.10 | 0.43 | 0.15 | 0.17 |
| | Res-FCN Model misfit (%) | 6.18 | 5.62 | 4.91 | 4.00 | 4.52 |
| | Res-FCN Data misfit (%) | 13.11 | 10.42 | 14.08 | 9.98 | 11.80 |

Case #1 is a ring with the relative permittivity of 1.5. It can be seen that at the selected five frequency points, compared with the results of VBIM, the results of Res-FCN are more uniform and the structure is clearer. At the same time, the model misfits are kept below 3%, while the inversion results of VBIM are about 5.5%. The data misfits of Res-FCN are also kept below 9%. For Case #2, there are two scatterers in DOI, a ring with the relative permittivity of 1.3 and a rectangle with the relative permittivity of 1.8. When the relative permittivity increases, compared with Case #1, the model misfits of VBIM increase by about 1%, and the outline of the rectangle becomes unclear. The model misfits of Res-FCN are still below than 5.5%, and the data misfits are below than 13%. In Case #3, the classical "Austria" model is employed as a test, which consists of two circles with the relative permittivity of 2 and an annulus with the relative permittivity of 1.5. It can be clearly observed that, at 1.7 GHz, the result obtained from VBIM, the reconstructed value of the small scatterer still fits well to the ground truth; as the frequency point increases to

1.9 GHz, the small scatterer cannot keep the basic circular shape. Meanwhile, the model misfits obtained from VBIM for Case #3 also increase to about 9%, while the proposed Res-FCN maintains good results across all five frequencies, and model misfits and data misfits are below 6.5% and 15%, respectively. Since VBIM is based on data misfit iteration, the data misfit of the traditional method is very low and the three cases are all below 1%.

Thus, it can be concluded that Res-FCN is more competent for the inversion problem with larger electrical sizes, compared with the conventional VBIM.

### C. Performance in the Entire Frequency Band

In this section, to further verify the inversion performance of the proposed Res-FCN in the entire frequency band, Cases #4-#6 with high permittivity scatterers are employed at 1001 frequency points with the step of 0.001 GHz from 1 GHz to 2 GHz, where 800 frequency points have not been trained. The corresponding model misfits and data misfits are shown in Figs. 5 and 6. Meanwhile, 11 inversion results uniformly taken according to the minimum and maximum value of the model misfit are shown in Fig. 7. Meanwhile, the corresponding model and data misfits of Cases #4-#6 with 11 inversion results at different frequency points are listed in Table III. Similar to the first example, Cases #4-#6 are more and more complicated, and the maximum value of the relative permittivity for scatterers are 6, 7, and 8, respectively.

| | Frequency (GHz) | 1.468 | 1.832 | 1.295 | 1.364 | 1.397 | 1.181 |
|---|---|---|---|---|---|---|---|
| Case #4 | Model misfit (%) | 1.34 | 3.74 | 4.31 | 6.09 | 8.13 | 10.09 |
| | Data misfit (%) | 7.77 | 21.51 | 36.92 | 50.05 | 22.14 | 71.65 |
| | Frequency (GHz) | 1.766 | 1.148 | 1.678 | 1.076 | 1.908 | |
| | Model misfit (%) | 13.35 | 17.97 | 20.91 | 25.80 | 26.18 | |
| | Data misfit (%) | 12.81 | 18.15 | 81.62 | 24.95 | 57.57 | |
| Case #5 | Frequency (GHz) | 1.018 | 1.241 | 1.338 | 1.565 | 1.753 | 1.907 |
| | Model misfit (%) | 1.63 | 2.04 | 3.36 | 5.40 | 7.03 | 9.57 |
| | Data misfit (%) | 21.41 | 21.26 | 23.58 | 42.55 | 37.40 | 80.79 |
| | Frequency (GHz) | 1.820 | 1.099 | 1.742 | 1.559 | 1.343 | |
| | Model misfit (%) | 13.22 | 20.85 | 23.23 | 26.11 | 27.67 | |
| | Data misfit (%) | 71.38 | 42.09 | 63.00 | 46.52 | 51.97 | |
| Case #6 | Frequency (GHz) | 1.650 | 1.603 | 1.476 | 1.173 | 1.332 | 1.813 |
| | Model misfit (%) | 7.28 | 8.40 | 9.07 | 10.37 | 11.70 | 12.39 |
| | Data misfit (%) | 51.69 | 66.27 | 53.73 | 58.61 | 59.79 | 85.16 |
| | Frequency (GHz) | 1.048 | 1.115 | 1.367 | 1.243 | 1.831 | |
| | Model misfit (%) | 16.77 | 18.81 | 20.10 | 24.39 | 27.30 | |
| | Data misfit (%) | 64.95 | 79.29 | 81.37 | 73.00 | 91.25 | |

Case #4 consists of two symmetrically placed circles with the relative permittivity of 4 and 6, respectively. In the entire frequency band, the smallest model misfit is at 1.468 GHz, and the corresponding model misfit and data misfit are 1.34% and 7.77%, respectively. The maximum model misfit is 26.18% at 1.908 GHz, and the corresponding data misfit is 57.57%. It can be seen that even at 1.908 GHz where the model misfit is the largest, the obtained inversion results can still maintain the shape of scatterers.

For Case #5, a rectangle is added to Case #4 as the third scatterer, and maximum value of the relative permittivity is increased to 7. It could be seen, when the frequency is at 1.018

GHz, the model misfit has the smallest value of 1.63%, and the corresponding data misfit is 21.41%. At 1.343 GHz, the model misfit reaches the maximum value of 27.67% and the corresponding data misfit is 51.97%. It can be seen that the proposed Res-FCN can obtain good inversion results in the entire frequency band based on these 11 inversion results which are from the smallest model misfit to the largest model misfit. However, with the relative permittivity increasing, the mismatch of pixels with high contrasts will bring large influence to the scattered field data, thus, the data misfits increase obviously.

In Case #6, an ellipse is added in the middle of two circles based on Case #5, so that the scatterers are getting closer. Meanwhile, the maximum value of the relative permittivity is increased to 8 to make the inversion more complicated. As shown in Fig. 5, the minimum value of the model misfit at 1.650 GHz is 7.28%, and the corresponding data misfit is 51.69%. At 1.831 GHz, the model misfit reaches the maximum value of 27.30%, and the data misfit is 91.25%. Based on the 11 inversion results which are from the minimum value to the maximum value of model misfit shown in Fig. 7, it can be seen that the shapes of scatterers are well reconstructed in the entire frequency band.

On the basis of the smallest model misfits of Case #4-#6, it can be seen that with the contrast increasing, a small model misfit will bring a huge change in the data misfit, this is because the scattered field data is more sensitive for the strong scatterer. Even so, as shown in Fig. 7, the proposed Res-FCN model exhibits good generalizability in the entire frequency band, and fits the huge nonlinearity caused by different frequency points in the range from 1 GHz to 2 GHz. Even for the untrained frequencies in a wide frequency band, Res-FCN can achieve good results. Meanwhile, even the results with the largest model misfits, the reconstructed permittivity distributions also can roughly match the ground truth.

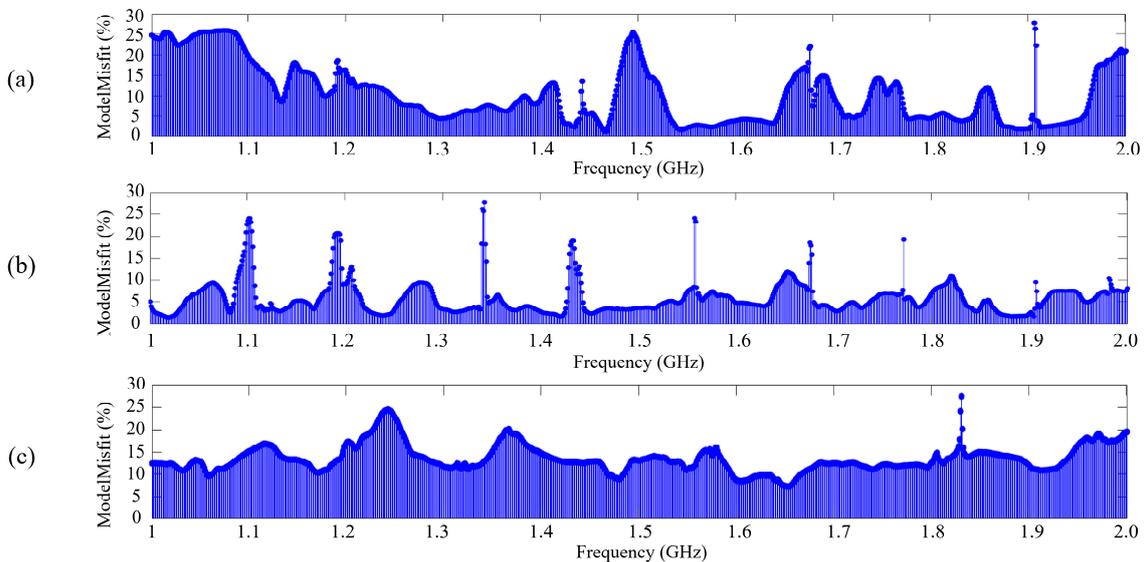

Fig. 5. Model misfit of the Case#4 to Case#6 across the entire frequency band with 0.001 GHz step shown in Fig. 7.

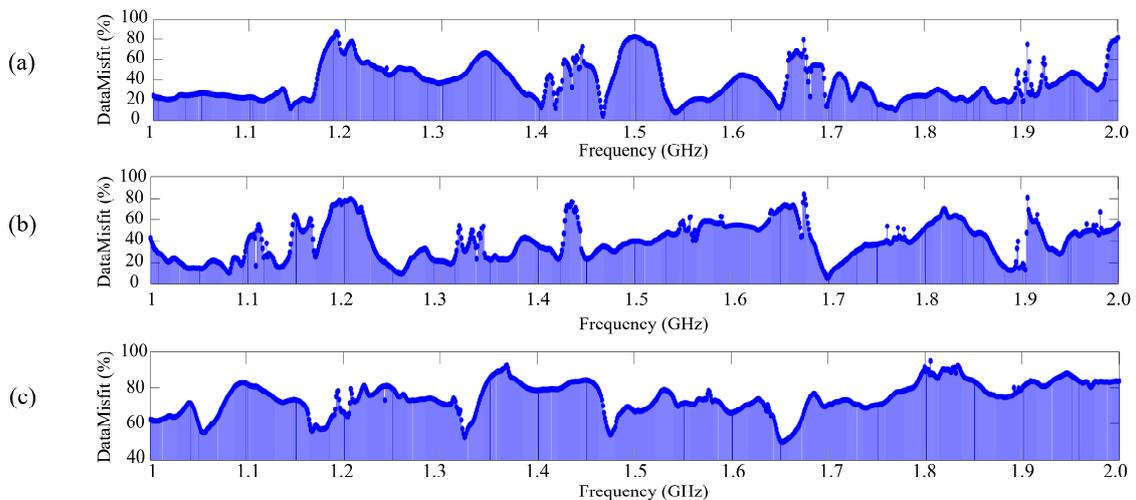

Fig. 6. Data misfit across of the Case #4 to Case #6 the entire frequency band with 0.001 GHz step shown in Fig. 7.

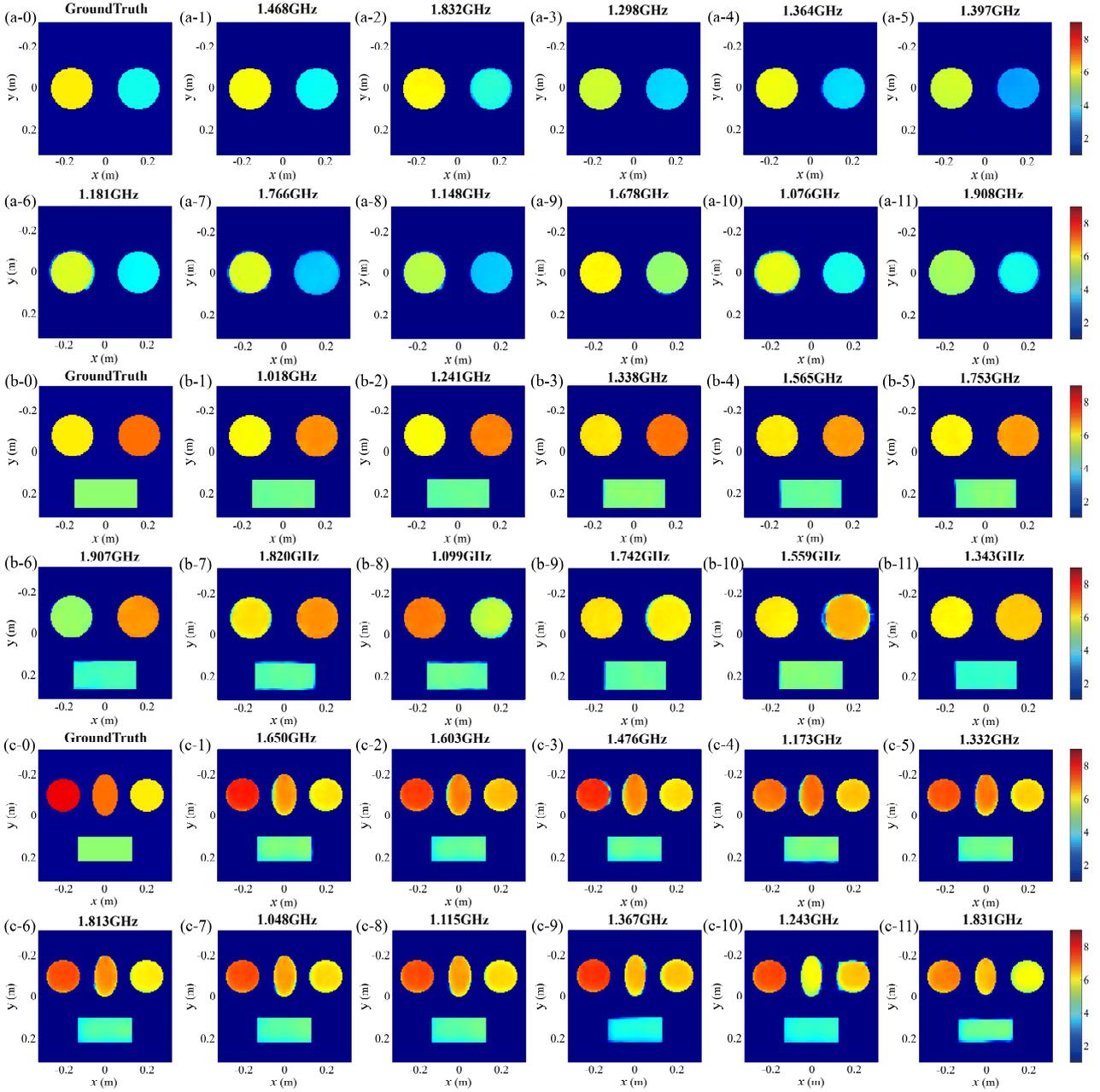

Fig. 7. Three cases are used to test the performance of the proposed model over the entire frequency band. (a-0), (b-0) and (c-0) are the ground truths of these three cases, respectively. (a-1), (b-1) and (c-1) are the results with the smallest model misfit in the entire frequency band, the model misfit from (a-1) to (a-11), (b-1) to (b-11) and (c-1) to (c-11) increases sequentially, and (a-11), (b-11) and (c-11) are the results with the largest model misfit in the entire frequency band, respectively

### D. Evaluation of Anti-Noise Ability of the Proposed Res-FCN

In this numerical example, to evaluate the anti-noise ability of the proposed Res-FCN, -30, -20 and -10 dB Gaussian white noise (i.e., the signal-to-noise ratio (SNR) is infinity, 30, 20 and 10 dB, respectively) are added. The inversion results of Case #7 at five randomly untrained frequency points in the frequency band are shown in Fig. 8, where the first row to the fourth row being the results of noise-free, -30 dB, -20 dB and -10 dB, respectively. In Case #7, the four contacting squares have the relative permittivity value of 5, 6, 7, and 8, respectively. The corresponding model and data misfits are listed in Table IV.

TABLE IV
THE MODEL MISFIT AND DATA MISFIT AND THEIR CORRESPONDING
FREQUENCIES SHOWN IN FIG. 8

| Frequency (GHz) | | 1.202 | 1.663 | 1.216 | 1.816 | 1.955 |
|---|---|---|---|---|---|---|
| NoiseFree | Modelmisfit (%) | 4.15 | 23.68 | 10.11 | 3.85 | 14.10 |
| | Datamisfit (%) | 28.89 | 77.98 | 64.91 | 21.52 | 73.21 |
| -30dB | Modelmisfit (%) | 4.18 | 23.77 | 10.78 | 3.78 | 14.61 |
| | Datamisfit (%) | 27.29 | 78.97 | 64.03 | 26.43 | 74.02 |
| -20dB | Modelmisfit (%) | 4.68 | 24.62 | 11.45 | 4.43 | 14.77 |
| | Datamisfit (%) | 25.56 | 74.08 | 60.90 | 24.78 | 74.22 |
| -10dB | Modelmisfit (%) | 8.17 | 28.14 | 18.55 | 4.71 | 19.44 |
| | Datamisfit (%) | 54.89 | 86.14 | 80.01 | 46.30 | 64.60 |

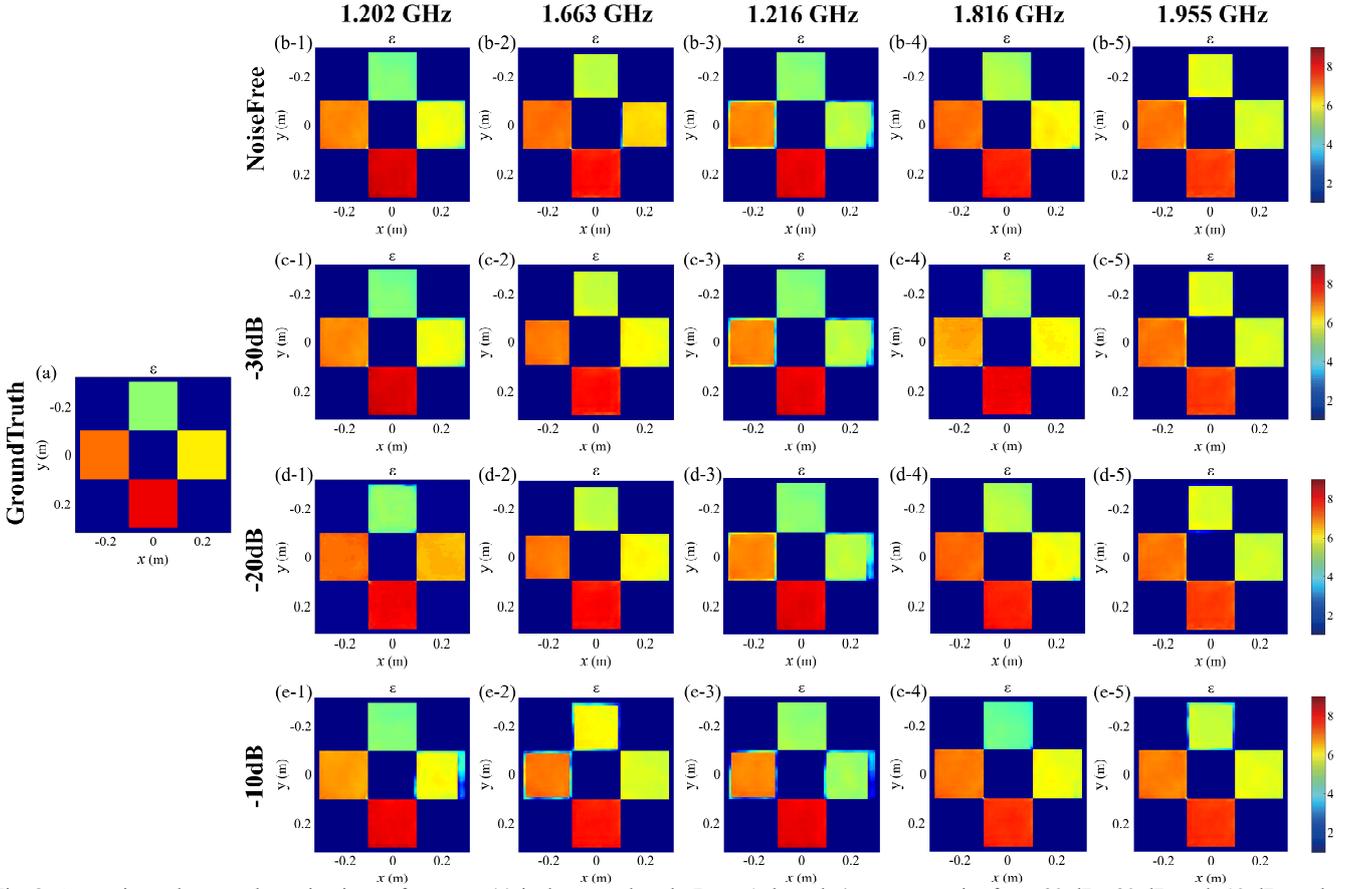

Fig. 8. A case is used to test the anti-noise performance. (a) is the ground truth. Rows 1 through 4 represent noise-free, -30 dB, -20 dB, and -10 dB results, respectively. Columns 1 through 5 are the results of five random frequencies under different noise levels.

It can be seen that when the noise-free result is good, such as 1.202 and 1.816 GHz, in the corresponding noisy environment, Res-FCN also can obtain good inversion results even with -10 dB noise. On the contrary, at 1.663 GHz, 1.216 GHz and 1.955GHz, the model misfits are larger in the noise-free environment, then as the noise increases, their model misfit increases more in the same noisy environment than those at 1.202 and 1.816 GHz. In general, the proposed Res-FCN can obtain good inversion results when the noise is less than -20 dB, although under the -10 dB noise environment, the inversion results are also in the acceptable range.

### E. Generalization Test of the Proposed Res-FCN

In the training dataset, the overlapping samples are not involved. Thus, in this example, two cases with overlapping scatterers are proposed to further verify the generalization of the Res-FCN, which are much different from the training dataset. As shown in Fig. 9, the first case consists of two tangential circles and the corresponding relative permittivity values are 6.4 and 3.0. Their model misfits are 32.25%, 24.48% and 34.76%, respectively. The second case consists of two tangential rectangles and the corresponding relative permittivity are 5.2 and 2.3. The boundary between the two scatterers is blurred, but the whole thing remains rectangular. The corresponding model misfits are 23.82%, 30.88% and 30.89% respectively. Though the relative permittivity of overlapping scatterers in the inversion results tend to be close to each other, this result is still good in the absence of overlapping samples in the training dataset.

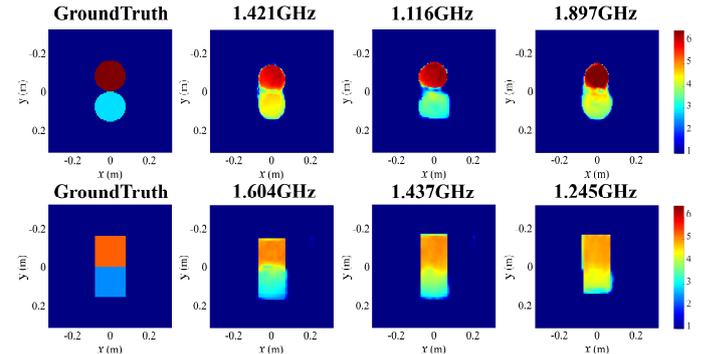

Fig.9 Two overlapping scatterers cases are used to test the generalization of the Res-FCN. The first column is ground truth. Columns 2-4 are inversion results under three random frequencies respectively.

### F. Experimental Test of the Proposed Res-FCN

In this example, to further verify the performance of the model, the measured data in [33] will be used to test the proposed model, which has a structure consisting of two symmetrical circular scatters. Moreover, to globally evaluate the proposed Res-FCN, the data in the range of 5 GHz to 6 GHz are employed, thus it is more nonlinear and difficult for the inversion problem compared to the lower frequency range. In the new model configuration, each pixel is represented by 1.5 mm and the size of the entire inversion area is 0.192 m×0.192 m. The experimental data need to be calibrated by the ratio of the measured incident field and the simulated incident field at the receiver opposite to the source [39]. As shown in Fig. 10, at 5 GHz and 6 GHz, the model misfits of inversion are 26.25% and 27.14%, respectively. From the inversion results, although the position of the cylinder is slightly offset, the Res-FCN accurately reconstructs the shape of scatterer and the relative dielectric constant value. The results fully demonstrate the performance and practicability of Res-FCN in practical measurements.

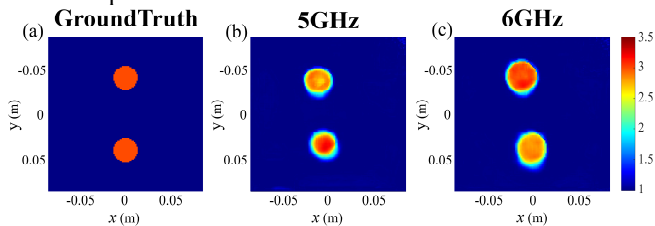

Fig. 10. An experimental case is used to test the Res-FCN. (a) is the ground truth. (b) is the inversion result at 5 GHz. (c) is the inversion result at 6 GHz.

## IV. CONCLUSION

To make the machine learning-based inversion method have more generalizability for the real engineering application, in this work, a Res-FCN is proposed to perform 2-D MWI for high contrast scatterers in a wide frequency band. By introducing the residual frameworks into FCN, the proposed Res-FCN is combined with the advantages of the Res-Net and FCN, which can fit strong nonlinearities from different frequency points and high contrast scatterers. Numerical examples verify the proposed Res-FCN can achieve good performance in the 2-D MWI with high contrast scatterers and anti-noise ability at an arbitrary frequency point in a wide frequency band. Even if there are no trained frequencies in the whole training frequency band, the proposed Res-FCN can reconstruct good results. The results of overlapped examples and measured data further prove the generalization of the proposed model.